\documentclass[runningheads]{llncs}
\usepackage{url}
\usepackage{caption}
\usepackage{subfigure}
\usepackage{adjustbox}
\usepackage{graphicx}
\usepackage{tabularx}
\usepackage[misc]{ifsym}
\usepackage{amsmath}
\usepackage{amssymb}
\usepackage{multirow}
\usepackage{bbding}
\usepackage{pifont}
\usepackage{wasysym}
\usepackage{utfsym}
\usepackage{fontawesome}
\usepackage{url}
\usepackage{mathtools}
\usepackage{color}
\usepackage{bm}
\usepackage{pifont}
\usepackage{multirow}
\usepackage{booktabs}

\usepackage{xcolor}
\definecolor{citecolor}{HTML}{00fa9a}
\definecolor{linkcolor}{HTML}{ED1C24}

\def\ie{\textit{i.e.}}
\def\eg{\textit{e.g. }}

\usepackage{hyperref}
\hypersetup{
	colorlinks=true,
	linkcolor=linkcolor,    
	urlcolor=linkcolor,
	citecolor=citecolor,
}

\begin{document}

\title{Tri-Plane Mamba: Efficiently Adapting Segment Anything Model for 3D Medical Images}
\titlerunning{TP Mamba}
%
\author{
Hualiang Wang\inst{1} \and 
Yiqun Lin\inst{1} \and 
Xinpeng Ding\inst{1} \and 
Xiaomeng Li\inst{1,2(\textrm{\Letter})}
}

\authorrunning{H. Wang et al.}
\institute{
The Hong Kong University of Science and Technology \\
\email{eexmli@ust.hk} \and
HKUST Shenzhen-Hong Kong Collaborative Innovation Research \\ Institute, Futian, Shenzhen}
\maketitle

\begin{abstract}
General networks for 3D medical image segmentation have recently undergone extensive exploration. Behind the exceptional performance of these networks lies a significant demand for a large volume of pixel-level annotated data, which is time-consuming and labor-intensive. The emergence of the Segment Anything Model (SAM) has enabled this model to achieve superior performance in 2D medical image segmentation tasks via parameter- and data-efficient feature adaptation. However, the introduction of additional depth channels in 3D medical images not only prevents the sharing of 2D pre-trained features but also results in a quadratic increase in the computational cost for adapting SAM.
To overcome these challenges, we present the \textbf{T}ri-\textbf{P}lane \textbf{M}amba (TP-Mamba) adapters tailored for the SAM, featuring two major innovations:  1) multi-scale 3D convolutional adapters, optimized for efficiently processing local depth-level information, 2)  a tri-plane mamba module, engineered to capture long-range depth-level representation without significantly increasing computational costs.
This approach achieves state-of-the-art performance in 3D CT organ segmentation tasks. Remarkably, this superior performance is maintained even with scarce training data. Specifically using only three CT training samples from the BTCV dataset, it surpasses conventional 3D segmentation networks, attaining a Dice score that is up to 12\% higher. %
The code is available at {\tt\small \url{https://github.com/xmed-lab/TP-Mamba}}.
\end{abstract}

\section{Introduction}

3D medical segmentation serves as a foundational process in medical analysis, crucial for accurate diagnosis diagnosis, treatment planning, and disease monitoring.
Courtesy of deep learning advancements, automatic segmentation methods have significantly advanced, achieving great success in tasks like tumor and organ segmentation~\cite{nnu,swinunetr,3duxnet,li2018h,wang2024towards,wang2023dhc,ctw,lin}.
Typically, the development of high-performance segmentation models for 3D medical images necessitates large quantities of high-quality annotated data, a process that is both time-consuming and labor-intensive.

Recently, the Segment Anything Model (SAM)~\cite{sam}, built upon vision transformer (ViT)~\cite{vit} and pre-trained on a billion-level dataset, has demonstrated remarkable general performance across various tasks in diverse domains and varying data scales~\cite{ma2024segment,wang2024samrs,li2023clip}, including the 2D medical field~\cite{samed,sammed2,sammed1}. 
Building on this success, fine-tuning the pre-trained SAM model with specific 3D medical datasets emerges as a promising method to improve segmentation performance, particularly when data is scarce.
However, adapting SAM for processing 3D medical data is non-trivial. Firstly, SAM is tailored for 2D images, thereby neglecting the crucial third-dimensional (depth) information present in medical images. Secondly, fine-tuning all parameters of SAM, given the limited 3D data, incurs high computational and time costs, and often leads to severe overfitting~\cite{pu2023empirical}. Moreover, this approach risks compromising the generalized knowledge acquired from the original large-scale data, potentially resulting in diminished performance\cite{ssf}.

\begin{figure}[t]    
  \centering    
  \subfigure[Performance comparison]
  { \label{fig:F11}
      \includegraphics[width=0.3\textwidth]{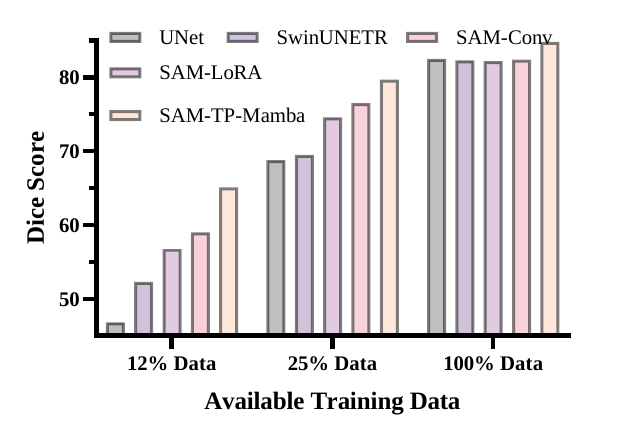}
  }
  \subfigure[Convergence rate]
  {\label{fig:F12}
      \includegraphics[width=0.3\textwidth]{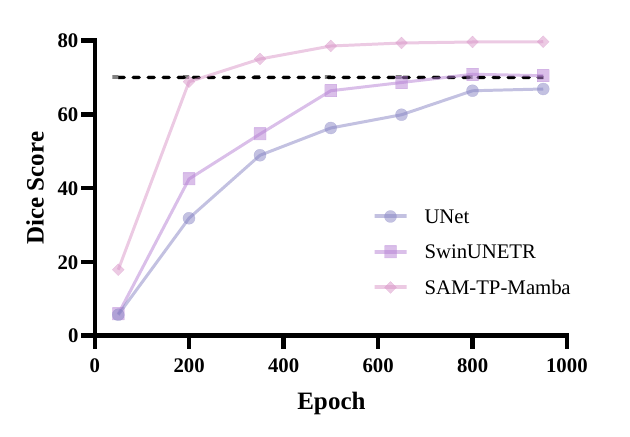}
  }
  \subfigure[Computation cost]
  {\label{fig:F13}
      \includegraphics[width=0.3\textwidth]{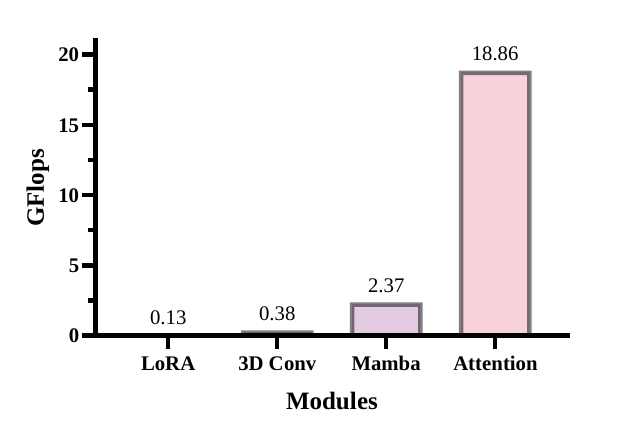}
  }
  \caption{ (a) Comparative performance analysis of TP-Mamba adapters against two conventional 3D segmentation networks and two adaptation methods, using 12\%, 25\%, and 100\% of the training data from the BTCV dataset. (b) Comparison of convergence rates between SAM-based adaptation and two conventional networks, using 25\% of the training data. (c) Increased GFlops per ViT block for four different adapters, given the input size $96 \times 96 \times 96$ and low-rank $r=96$.}    
  \label{fig:F1}   
\end{figure}

To address the above problems, SAMed~\cite{samed} first treats slices at different depths as separate batch images and transfers the 3D data to 2D slices.
Furthermore, it freezes the parameters of SAM, and only tunes the lightweight Low-Rank adapters (LoRA)~\cite{hu2021lora} integrated into SAM, maintaining the learned knowledge while reducing the fine-tuning cost.
Although intuitive, SAMed only captures the features of 2D slices, and neglects the inter-slice (\ie,~the depth) correlations.
To bridge this gap, MA-SAM~\cite{masam} and 3DSAM-Adapter~\cite{samadapter} both introduce 3D convolutional adapters for capturing depth-wise information. 
Equipped with 3D convolutional layers, the original SAM can capture translational invariances and recognize local 3D features.
However, due to the intrinsic locality of convolution operations, they exhibit limitations in modeling long-range dependencies.

Considering the success of self-attention (SA) in capturing global relationships, a naive solution is to divide the 3D data into multiple patches and subsequently deploy SA modules as low-rank adapters to learn the long-range correlations existing among these patches.
However, the SA requires quadratic complexity in terms of the number of patches, resulting in expensive computational overhead when addressing 3D segmentation tasks.
As shown in Fig.~\ref{fig:F13}, given the input with the size of $96 \times 96 \times 96$ and low rank $r=96$, the increased GFlops per block in a ViT-B (consisting of 12 blocks in total) will amount to \textit{18.86}, nearly \textit{\textbf{145}} times greater than that of the standard LoRA.
Recent advancements in state space models, especially Mamba~\cite{mamba}, present a promising solution for learning global features with significantly lower computational costs.
There are also some researchers adapting Mamba to the medical domain,~\eg, U-Mamba~\cite{umamba}, to reduce the computation cost of vision transformers.
These methods simply flatten 3D images into 2D sequences and directly use mamba to model the causal relations between the sequences,~\ie, the current elements can only interact with any of the previously scanned samples through a compressed hidden state.
However, due to the non-casual nature of 3D medical data, naively applying this approach to flattened sequences cannot learn the relations between unscanned patches, resulting in limited receptive fields.

In this paper, we introduce a novel tri-plane mamba (TP-Mamba) adapter and incorporate it into SAM to capture both local and global 3D non-casual information of medical images in a parameter-efficient way.

The proposed TP-Mamba adapter has two key components.
Specifically, a mixture of four parallel 3D convolutional layers with different dilation rates is used to capture the 3D multi-scale local depth-wise features.
Then, a tri-plane scan module is designed to structurally scan the 3D features along three 2D planes (depth-height, depth-width, and height-width planes), as shown in Fig.~\ref{fig:F23}.
Different from the single-direction scanning in the original Mamba, our module can ensure that each element in a 3D feature is attended by other elements from different directions.

During the training, we only fine-tune the proposed TP-Mamba adapters while freezing all other parameters of SAM. 
Overall, the TP-Mamba adapters, proficient in capturing multi-scale and local-global feature representations for 3D medical data, also ensure \textbf{parameter- and data-efficient adaptation}, thereby facilitating \textbf{rapid convergence}, as shown in Fig.~\ref{fig:F11} and Fig. ~\ref{fig:F12}.

Comprehensive experiments conducted on organ segmentation tasks with different data scales demonstrate that TP-Mamba adapters achieve state-of-the-art performance against ten baselines consisting of Transformer-based, convolutional-based, Mamba-based networks, and SAM-based fine-tuning algorithms.

\section{Methodology}
\subsection{Architecture Overview}

\begin{figure}[t!]    
  \centering    
  \subfigure[Overview of the pre-trained SAM with adapters on 3D segmentation. ]
  { \label{fig:F21}
      \includegraphics[width=0.85\textwidth]{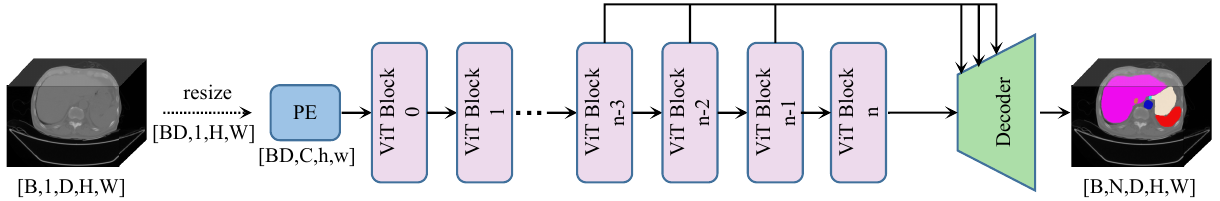}
  }
  \subfigure[ViT block]
  {\label{fig:F22}
      \includegraphics[width=0.2\textwidth]{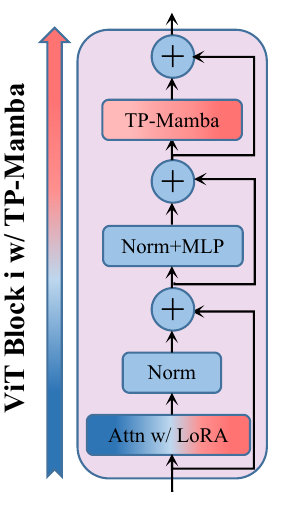}
  }
  \subfigure[Architectural details of the proposed Tri-plane mamba]
  {\label{fig:F23}
      \includegraphics[width=0.65\textwidth]{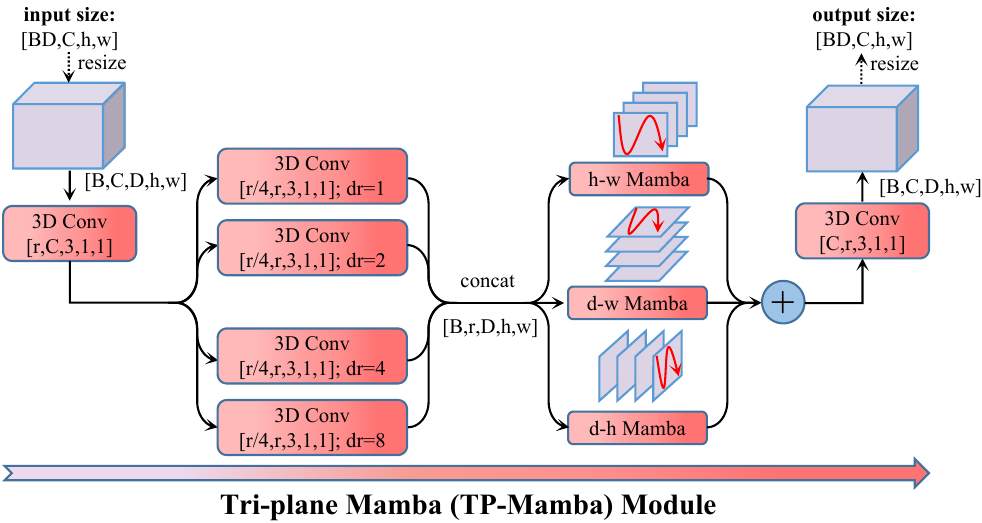}
  }
\caption{(a) Overview of the pre-trained SAM with adapters on 3D segmentation tasks. Given the 3D images, the patch embedding layer and 12 ViT blocks from the pre-trained SAM will extract 3D features. A decoder is used to predict the segmentation maps via four feature maps from the last four blocks. (b) The details of the ViT block equipped with LoRA and the proposed TP-Mamba adapters. (c) Architectural details of the proposed TP-Mamba adapters.} 
\label{fig:pipeline}
\end{figure}

The architecture of the proposed tri-plane mamba adapters for 3D-SAM is sketched in Fig.~\ref{fig:F21}. The input 3D image batch $\bm{\mathrm{X}} \in \mathbb{R}^{B \times 1 \times D \times H \times W}$ is first re-shaped to $BD \times 1 \times H \times W$ and fed into the patch embedding module of the pre-trained SAM model to extract the patch embeddings $\bm{\mathrm{F}}^0 \in \mathbb{R}^{BD \times C \times h \times w}$, where $C$ is the number of feature dimensions, $D$ is the number of slices (depths) per 3D image and $(H, W)/(h, w)$ is the size of original/down-sampled images/feature maps. Next, the sequential ViT blocks of pre-trained SAM are used to extract features as follows:
\begin{equation}
    \bm{\mathrm{F}}^i = \psi^i(\bm{\mathrm{F}}^{i-1}),
\end{equation}
where $\bm{\mathrm{F}}^i \in \mathbb{R}^{BD \times C \times h \times w}$ indicates the feature embeddings extracted from $i$-th ViT blcok of the SAM, where $i \in [1,n]$. $\psi^i$ is the $i$-th ViT block with the proposed tri-plane mamba adapters and low-rank adapters (LoRA)~\cite{hu2021lora}. 

The structure of ViT block $\psi$ is illustrated as Fig.~\ref{fig:F22}. Features will be fed into a multi-head self-attention module where LoRA is inserted into it~\cite{hu2021lora}. Next, a normalization function (instantiated as the layer normalization), multi-layer perception (MLP), and another normalization function are sequentially deployed to extract features. Finally, the proposed tri-plane mamba module is used to capture both local and global 3D features as the output of ViT block.

After encoding image features, a decoder model is designed to predict the final segmentation maps $\bm{\mathrm{P}} \in \mathbb{R}^{B \times K \times D \times H \times W}$, where $K$ is the number of classes. Finally, the Dice and cross-entropy loss function~\cite{sudre2017generalised} is utilized to calculate the training loss for each batch data:
\begin{equation}
    loss = \mathcal{L}_{DiceCE}(\bm{\mathrm{P}}, \bm{\mathrm{Y}}),
\end{equation}
where $\bm{\mathrm{Y}}$ is the ground-truth maps.

Next, we will elaborate on the details of the tri-plane mamba module, and decoder model.

\subsection{Tri-plane Mamba Module}
\begin{figure}[t!] 
\centering 
\includegraphics[width=0.9\textwidth]{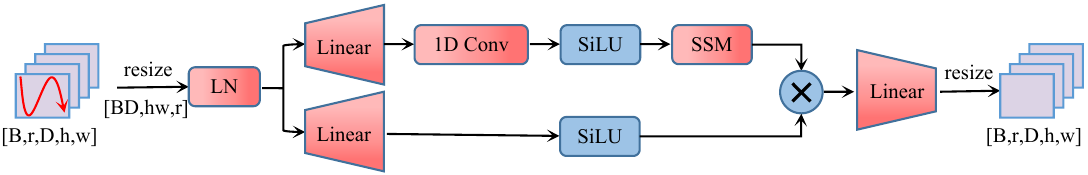} 
\caption{The illustration of the SSM-based mamba~\cite{mamba} block $\phi^{hw}$ in Eqn.~\ref{eq3}, exemplified using height-width plane scanning. } 
\label{fig:3}
\end{figure}

To parameter-efficient adapt 3D features for a 2D pre-trained SAM model, we propose to employ a tri-plane mamba module to aggregate 3D features from three planes. Specifically, given a feature tensor $\bm{\mathrm{F}} \in \mathbb{R}^{BD \times C \times h \times w}$ from the last layer of the ViT block, we first reshape it as the standard 3D format, (batch, features, depth, height, width). Then a 3D convolution layer with kernel size $r \times C \times 3 \times 1 \times 1$ is used to reduce the feature dimension from $C$ to $r$, keeping the adapters parameter-efficient. Considering that the main body of the ViT block ignores the learning of depth-level features, four 3D convolution layers with kernel size $\frac{r}{4} \times r \times 3 \times 1 \times 1$
and dilation rates [1,2,4,8] are deployed to extract multi-scale local representations parallelly. Then feature tensors from four layers are concentrated together, denoted as $\bm{\mathrm{F}} \in \mathbb{R}^{B \times r \times D \times h \times w}$.

Next, $\bm{\mathrm{F}}$ are split and flattened from three planes (\ie, height-width, depth-height, and depth-width planes). The processed long-sequential feature tensors are denoted as $\bm{\mathrm{F}}^{hw} \in \mathbb{R}^{BD \times hw \times r}$, $\bm{\mathrm{F}}^{dh} \in \mathbb{R}^{Bw \times Dh \times r}$ and $\bm{\mathrm{F}}^{dw} \in \mathbb{R}^{Bh \times Dw \times r}$, respectively, as shown in Fig.~\ref{fig:3}. Next, three Mamba blocks~\cite{mamba} $\phi^{hw}$, $\phi^{dw}$ and $\phi^{dh}$ are utilized for highly efficient sequential feature learning with long-range correlations from tri-plane features, respectively. After the Mamba blocks, the features are added together. 
This process can be defined as:
\begin{equation}
\label{eq3}
\begin{aligned}
\bm{\mathrm{F}} &= \phi^{hw}(\bm{\mathrm{F}}^{hw}) + \phi^{dw}(\bm{\mathrm{F}}^{dw}) + \phi^{dh}(\bm{\mathrm{F}}^{dh}).
\end{aligned}
\end{equation}

The detailed module pipeline of the mamba block is illustrated in Fig.~\ref{fig:3}. Finally, another 3D convolutional layer with the kernel size of $C \times r \times 3 \times 1 \times 1$ is used to increase the feature dimension to $C$.

\textbf{Decoder} The feature maps from the last four ViT blocks are concentrated together and fed into a series of 3D convolutional, instance normalization, GELU activation~\cite{hendrycks2016gaussian}, and upsampling layers to recover the resolutions of features as the original one. Finally, a convolutional layer is implemented to predict the segmentation maps $\bm{\mathrm{P}} \in \mathbb{R}^{B \times K \times D \times H \times W}$ via the upsampled feature maps.

\subsection{Implementation Details}
\textbf{Dataset:}
We conducted comprehensive evaluations of our method on the Beyond the Cranial Vault (BTCV) dataset with the setting of employing varying quantities of training data. The BTCV dataset, as referenced in the study by~\cite{btcv}, comprises 30 CT volumes annotated manually for 13 abdominal organs.
We split the dataset into 24 training and 6 testing cases for the full training data utilization scenario. In scenarios with reduced training data, specifically 25\% and 12\%, we randomly selected 6 and 3 training cases five times respectively. The performance metrics reported are the averages of these five runs, ensuring a robust evaluation of our method under varied training data conditions.

\noindent\textbf{Data Processing:} The intensity values of each CT scan in BTCV dataset were truncated within the range of [-200, 250] Hounsfield Units (HU) for observing abdominal organs. The HU values are then normalized to the range of [0, 1] using min-max normalization. Next, the data is resampled to 1.0 mm isotropic spacing during training and inference, using the input image size of 96×96×96 (depth, height, and width) and batch size 1. The sliding-window inference strategy is employed when evaluating the test set's performance.

\noindent\textbf{Training and inference Details:} We use pre-trained SAM based on ViT-B-16\footnote{\href{https://huggingface.co/timm/samvit\_base\_patch16.sa1b}{https://huggingface.co/timm/samvit\_base\_patch16.sa1b}} as our backbone network. 
For all experiments, the AdamW optimizer~\cite{adamw} is applied, with the learning rate gradually reduced from $0.0002$ to $0$ as the 1,000 training epochs. 
Training data augmentation includes spacing, random cropping, flipping, and contrast adjustments.
All experiments are implemented with Pytorch~\cite{imambi2021pytorch}. 
We assess model performance using the Dice similarity coefficient score at 1.0 mm tolerance for volumetric accuracy. 
The uniform framework serves as a standard testbed for evaluating all models, ensuring neutrality and fairness. This approach deliberately avoids favoring any model based on variables such as input size, spacing, data augmentations or evaluation techniques.

\noindent\textbf{Compared methods:} We select eight renowned models designed for medical volumetric segmentation tasks to benchmark our approach, including nnUNet~\cite{nnu}, VNet~\cite{vnet}, UNETR~\cite{unetr}, SwinUNETR~\cite{swinunetr}, TransBTS~\cite{wang2021transbts}, 3D-UX-Net~\cite{3duxnet}, MedNeXt~\cite{roy2023mednext}, and U-Mamba~\cite{umamba}.
In addition, we also evaluate two adapter algorithms for SAM: SAMed~\cite{samed} which utilizes LoRA adapters, and MA-SAM~\cite{masam}, employing 3D convolution adapters.

\section{Experiments}

\begin{table}[t!]
\centering
\caption{Segmentation results on BTCV dataset. The red and blue numbers respectively indicate the best and second-best Dice score.\\
\label{tab:results1}
}
\setlength{\tabcolsep}{5pt}
\begin{adjustbox}{width=1.0\textwidth}
\begin{tabular}{c||ccccccccccccc||c}
\toprule
Method & \rotatebox{45}{Sple}  & \rotatebox{45}{Rkid} & \rotatebox{45}{Lkid} & \rotatebox{45}{Gall} & \rotatebox{45}{Eso} & \rotatebox{45}{Liver} & \rotatebox{45}{Sto} & \rotatebox{45}{Aorta} & \rotatebox{45}{IVC} & \rotatebox{45}{Veins} & \rotatebox{45}{Panc} & \rotatebox{45}{Rad} & \rotatebox{45}{Lad} & Avg \\ 
\midrule
\multicolumn{15}{c}{100\% of The Training Data (24 samples)} \\
\midrule
nnUNet & 96.4 &	94.7 	&94.9 	&74.8 &	75.5 &	\textcolor{blue}{97.1} 	&79.4 &	90.8 	&84.3 	& \textcolor{blue}{77.7} &	79.4 	&64.4 	&63.3 &	82.5   \\
VNet & 92.8 &	91.7 &	93.0 &	70.5 &	72.2 &	96.1 &	81.0 &	88.7 &	83.5 &	68.1 &	77.9 &	64.2 &	60.1 &	80.0  \\
UNetr & 91.5 &	92.8 	&91.2 &	60.2 &	72.0 &	93.0 	&80.1 	&87.5 	&80.2 	&67.8 &	71.8 	&61.6 	&57.4 	&77.5  \\
TransBTS & 94.1 &	93.9 &	93.7 	&75.9 &	69.2 &	96.5 &	\textcolor{blue}{82.5} 	&88.8 	&85.0 &	73.1 	&77.1 	& \textcolor{blue}{68.1} &	55.0 	&81.0  \\
SwinUNetr & 95.7 &	94.4 &	94.1 &	71.5 &	75.7 &	96.9 &	78.8 	&90.1 &	\textcolor{blue}{85.7} 	&74.5 	& \textcolor{blue}{80.7} &	66.7 &	64.6 &	82.3  \\
3D-UX-Net & 95.3 &	94.2 &	93.6 &	71.9 &	71.6 &	96.5 	&82.3 	&90.4 	&84.8 &	71.3 	&76.8 	&67.5 &	\textcolor{blue}{66.4} &	81.8  \\
MedNeXt  & \textcolor{blue}{96.5} &	\textcolor{red}{95.0} &	\textcolor{red}{95.0} &	71.5 &	\textcolor{red}{77.2} &	\textcolor{red}{97.2} &	81.7 &	\textcolor{blue}{91.0} &	85.4 	&77.1 	&79.3 &	68.0 &	65.9 &	\textcolor{blue}{83.1}  \\
U-Mamba & 95.8 &	94.5 &	94.6 &	73.8 &	75.1 &	96.9 &	81.7 &	90.8 &	84.7 &	75.7 	&79.3 	&64.4 &	57.7 &	81.9  \\
\midrule
SAMed & 93.4 &	94.7 &	\textcolor{blue}{94.9} &	\textcolor{blue}{76.2} &	75.5 &	96.7 &	78.7 &	90.6 &	84.1 &	\textcolor{red}{78.1} &	77.9 	&64.1 &63.9 &	82.2  \\
MA-SAM & 96.3 &	\textcolor{blue}{94.9} &	94.8 &	73.3 &	74.8 &	\textcolor{blue}{97.1} &	82.4 &	90.8 &	83.6 &	76.5 &	79.8 	&63.8 	&63.7 	&82.4 \\
\midrule
TP-Mamba & \textcolor{red}{96.6} &	\textcolor{red}{95.0} 	& \textcolor{blue}{94.9} &	\textcolor{red}{79.6} 	& \textcolor{blue}{76.3} 	& \textcolor{red}{97.2} &	\textcolor{red}{86.6} &	\textcolor{red}{91.5} &	\textcolor{red}{87.1} &	77.0 	& \textcolor{red}{85.1} &	\textcolor{red}{68.3} 	& \textcolor{red}{67.1} &	\textcolor{red}{84.8} \\
\midrule
\multicolumn{15}{c}{25\% of The Training Data (6 samples)} \\
\midrule
nnUNet & 84.2 &	90.8 &	91.0 &	42.6 &	45.4 	&94.5 &	52.4 &	81.4 &	78.8& 	61.8 &	63.4 	&57.7 &	50.0 &	68.8  \\
VNet & 74.1 &	85.0 &	81.5 &	44.7 &	0.0 &	86.8 &	46.7 	&77.1 &	76.4 &	0.2 	&46.7 &	0.0 	&0.0 &	47.6  \\
UNetr &78.1 &	86.2 	&80.1 &	0.0 	&0.0 	&90.6 &	42.1 	&78.6 &	64.4 &	40.4 &	32.1 &	0.0 &	0.0 	&45.6 \\
TransBTS & 62.3 &	80.4 &	73.1 &	40.7 	&0.0 &	94.7 	&25.0 &	73.5 &	62.1 &	46.4 	&27.9 &	0.0 	&0.0 &	45.1 \\
SwinUNetr &83.0 &	89.8 &	89.2 	&36.1& 	55.2 &	90.3 	&59.2 	&83.6 &	77.2 &	65.2 	&60.8 	&60.4 &	54.1 &	69.5 \\
3D-UX-Net &76.4 &	87.8 &	88.3 &	53.1 &	\textcolor{blue}{64.0} 	&94.1 	&47.8 &	85.5 &	81.2 &	66.7 	&59.0 &	58.9 &	56.6 &	70.7 \\
MedNeXt  & 81.3 &	91.4 &	90.1 	&52.6& 	57.1 &	92.8 	&50.8 &	84.4 	&78.7 	&68.4 	&64.7 &	\textcolor{red}{63.3} &	\textcolor{red}{59.1}	&71.9 \\
U-Mamba & 87.2 &	87.6 &	87.4 &	46.6 &	63.5 &	92.8 &	42.6& 	82.9& 	74.5 	&66.7& 	64.9 &	48.5 	&46.2& 	68.6   \\
\midrule
SAMed & 90.8 &	89.2 &	90.0 &	62.1 &	\textcolor{red}{64.4} &	94.4 &	\textcolor{blue}{64.5} &	\textcolor{blue}{87.0} &	78.8 &	69.2 &	70.0 	&56.6 	&52.8 &	74.6  \\
MA-SAM & \textcolor{blue}{90.9} & 	\textcolor{blue}{92.5} &	\textcolor{blue}{92.7} &	\textcolor{blue}{74.2} 	&63.8 	& \textcolor{blue}{95.9} 	&63.6 &	86.6 &	\textcolor{blue}{81.8} &	\textcolor{blue}{71.1} &	\textcolor{blue}{70.6} 	& 57.2 &	53.4 &	\textcolor{blue}{76.5}  \\
\midrule
TP-Mamba & \textcolor{red}{94.5} &	\textcolor{red}{93.4} &	\textcolor{red}{93.7} &	\textcolor{red}{75.8} &	63.1 &	\textcolor{red}{96.4} 	& \textcolor{red}{77.2} & 	\textcolor{red}{87.2} &	\textcolor{red}{85.2} &	\textcolor{red}{73.2} 	& \textcolor{red}{78.9} &	\textcolor{blue}{60.9} 	& \textcolor{blue}{56.7} &	\textcolor{red}{79.7}  \\
\midrule
\multicolumn{15}{c}{12\% of The Training Data (3 samples)} \\
\midrule
nnUNet & \textcolor{blue}{80.7} & 	82.8& 	57.3 &	1.9 &	35.3 &	91.2 	&35.4 &	78.4 &	73.6 &	27.1 &	32.4 &	2.4 &	9.5 	&46.8  \\
VNet & 1.9 &	54.6 &	0.0 &	0.0 &	0.0 &	81.7 	&25.1 &	36.6 	&1.9 &	0.0 &	0.0 	&0.0 &	0.0 	&15.5 \\
UNetr & 72.4 &	56.7 &	0.0 &	0.0 &	0.0 &	87.9 	&26.4 &	26.7 	&0.0 	&0.0 	&0.5& 	0.0 &	0.0 	&20.8  \\
TransBTS & 29.3 &	47.1 &	21.6 &	0.0 &	0.0 &	90.7 	&11.2 &	17.0 &	0.0 	&23.6 &	8.6 &	0.0 &	0.0 	&19.2 \\
SwinUNetr & 64.3 &	79.5 &	58.8 	&7.5 &	45.6 	&90.7 	&37.7 &	79.4 	&69.6 &	49.4 &	26.4 &	\textcolor{blue}{38.6} &	32.8 	&52.3 \\
3D-UX-Net &72.0 &	80.5 &	67.2 	&26.1& 	43.8 &	89.5 	&35.4 &	80.9 	& \textcolor{blue}{74.3} & 	45.6 &	17.2 &	\textcolor{red}{42.3} &	16.6 	&53.2 \\
MedNeXt  & 42.1 &	76.3 &	62.0& 	\textcolor{blue}{36.1} &	\textcolor{red}{55.6} &	87.5 	&36.0 &	80.1 	&70.0 &	58.3 &	37.4 &	29.9 &	24.5 	&53.5 \\
U-Mamba & 66.9 &	72.8 &	58.9 &	20.9 &	51.5 	&90.7 	&36.2 &	78.9 &	71.1 &	42.0 &	13.6 	&36.0 &	10.2 	&50.0 \\
\midrule
SAMed & 49.6 &	78.0 &	68.0& 	33.8 &	50.9& 	89.5 	&41.0& 	82.3 	&71.3 &	\textcolor{red}{62.3} &	\textcolor{blue}{42.4} &	29.9 &	25.5 	&55.7 \\
MA-SAM & 64.1 &	\textcolor{blue}{84.0} & 	\textcolor{blue}{75.5} &	33.2 &	\textcolor{blue}{55.4} 	& \textcolor{blue}{92.6} 	& \textcolor{blue}{45.2} &	\textcolor{blue}{84.1} 	&72.8 &	58.8 &	37.6 	&30.1 &	\textcolor{blue}{33.1} 	& \textcolor{blue}{59.0}  \\
\midrule
TP-Mamba & \textcolor{red}{88.6} &	\textcolor{red}{90.7} &	\textcolor{red}{81.1} &	\textcolor{red}{42.8} & 	52.3 &	\textcolor{red}{95.3} 	& \textcolor{red}{59.4} &	\textcolor{red}{85.8} 	& \textcolor{red}{80.5} &	\textcolor{blue}{59.8} &	\textcolor{red}{45.9} &	37.7 	& \textcolor{red}{35.3} 	& \textcolor{red}{65.8} \\
\bottomrule
\end{tabular}
\end{adjustbox}
\end{table}

\subsection{Evaluation on BTCV Dataset}
Table.~\ref{tab:results1} presents a detailed comparison of the performance between our method and ten baseline networks on the BTCV dataset, using varying proportions of training data: 100\%, 25\%, and 12\% respectively.

Table.~\ref{tab:results1} demonstrates that: \ding{182} \textbf{our algorithm consistently outperforms others across various settings.} Specifically, compared to other 3D segmentation networks, our method surpasses the best-performing network by margins of 1.7\%, 7.8\%, and 12.3\% in average Dice score. Additionally, when compared to adapter algorithms, our method maintains a lead of at least 2.4\%, 3.2\%, and 6.8\% in average Dice score; \ding{183} \textbf{Our algorithm exhibits notable data efficiency.} Even when the training data is reduced from 100\% to 25\% and 12\%, our method maintains a relatively stable performance level, in contrast to the significant performance deterioration observed in other comparative methods.

\begin{figure}[t]    
  \centering    
  \subfigure[12\% training data]
  { \label{fig:F41}
      \includegraphics[width=0.3\textwidth]{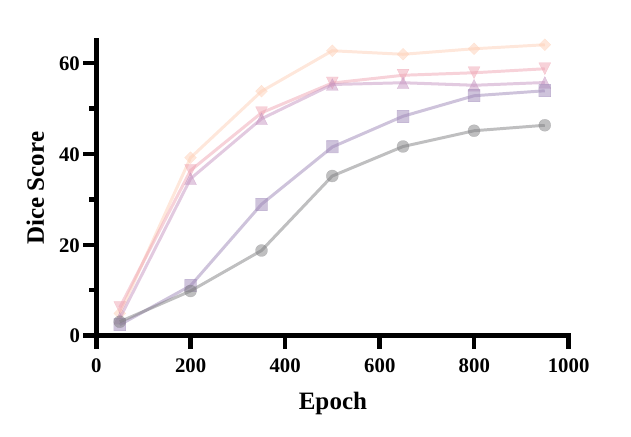}
  }
  \subfigure[25\% training data]
  {\label{fig:F42}
      \includegraphics[width=0.3\textwidth]{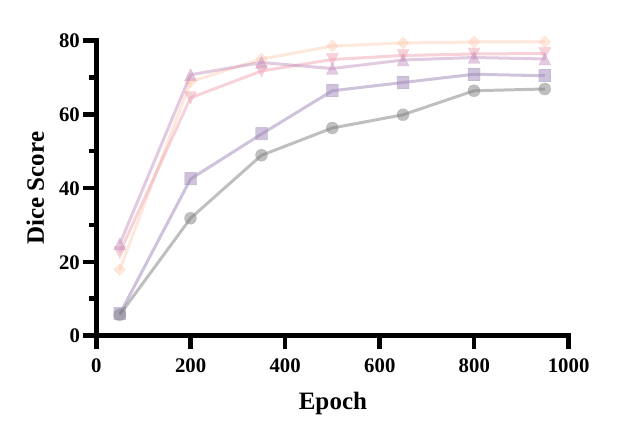}
  }
  \subfigure[100\% training data]
  {\label{fig:F43}
      \includegraphics[width=0.3\textwidth]{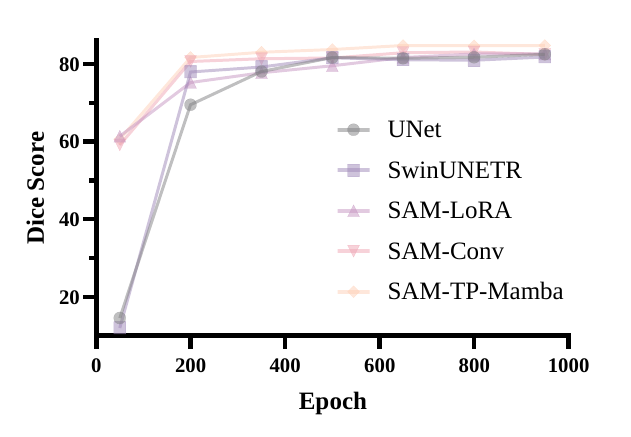}
  }
  \caption{Comparison of convergence rates between our method and two conventional
networks and two adapter algorithms on three settings.}    
  \label{fig:F4}     
\end{figure}

Fig.~\ref{fig:F4} presents a comparative analysis of the Dice score curves for different networks during the training process. This illustration clearly shows that our algorithm \textbf{converges more rapidly}, a benefit attributable to the \textbf{parameter-efficient adapters} and the extensive prior knowledge encapsulated in the pre-trained SAM. Specifically, when trained with only 12\% of the data, our algorithm achieves performance comparable to traditional segmentation networks by around the 400th epoch. Furthermore, with 25\% and 100\% of the training data, it surpasses the performance of segmentation networks trained for 1000 epochs in just around 200 epochs.

\begin{table}[h]
\centering
\begin{minipage}{0.31\linewidth}
\centering
\caption{Ablation on 3D conv adapters}
\label{tab:T21}
\begin{tabular}{c|c} 
\toprule
3D Conv & Dice  \\
\midrule 
conv (k=3) & 75.1  \\
MS conv (k=3) & 78.9  \\
conv (k=5) & 75.9   \\
MS conv (k=5) & 79.0 \\
\bottomrule
\end{tabular} 
\end{minipage}
\begin{minipage}{0.31\linewidth}  
\centering
\caption{Ablation on the scanning of mamba}
\label{tab:T22}
\begin{tabular}{c|c} 
\toprule
Mamba & Dice  \\
\midrule 
Volume & 73.9  \\
hw-plane & 74.4  \\
dw-plane & 76.5  \\
tri-plane & 78.7   \\
\bottomrule
\end{tabular} 
\end{minipage}
\begin{minipage}{0.31\linewidth}  
\centering
\caption{ Ablation on the value of low rank $r$}
\label{tab:T23}
\begin{tabular}{c|c} 
\toprule
$r$ & Dice  \\
\midrule 
$r=24$ & 78.3  \\
$r=48$ & 78.0  \\
$r=96$ & 78.9  \\
$r=192$ & 78.9   \\
\bottomrule
\end{tabular} 
\end{minipage}
\end{table}

\subsection{Ablation Analysis}
All models of ablation study are trained with 25\% of the training data. 

\noindent\textbf{Effectiveness of the multi-scale 3D convolution module.} 
To validate the effectiveness of this module, we conducted four sets of experiments. Firstly, we replaced the multi-scale (MS) 3D convolution layers with kernel size $\frac{r}{4} \times r \times 3 \times 1 \times 1$ shown in Fig.~\ref{fig:F23} with a single 3D convolution layer with kernel size $r \times r \times 3 \times 1 \times 1$. Secondly, we increased the kernel size along the depth dimension from $3$ to $5$. Table.~\ref{tab:T21} shows that the MS design significantly enhances performance, benefiting from the expansion of the model's local receptive field.

\noindent\textbf{Effectiveness of the scanning strategy.} 
Table.~\ref{tab:T22} shows that scanning from depth-width planes surpasses both volume flattening and scanning from height-width planes, primarily due to its more effective capture of depth-level and 3D structural information. Furthermore, ensemble scanning across all three planes (height-width, depth-width, and depth-height) further boosts performance

\noindent\textbf{Effectiveness of the low rank $r$.} As shown in Table.~\ref{tab:T23}, our algorithm maintains a consistent performance across different values of $r$, presenting strong robustness without significant fluctuations in performance.

\section{Conclusion}

In conclusion, this study presents a new advanced network in 3D medical image segmentation, which adapts the Segment Anything Model (SAM) via the proposed novel tri-plane mamba (TP-Mamba) adapters. To be specific, TP-Mamba adapters feature multi-scale 3D convolution layers and a tri-plane mamba module. These adaptations effectively enable local depth-level information processing and capture long-range depth-level features without significantly increasing computational demands. As a result, our approach not only mitigates the reliance on extensively annotated datasets but also addresses the computational inefficiencies of SAM's adaptation on 3D medical image segmentation.
Our method demonstrates its superiority in 3D CT organ segmentation, maintaining exceptional performance even when trained with limited data.

\noindent\textbf{Limitations and future work.} This work focuses solely on customizing the image encoder of SAM for 3D medical image segmentation, while overlooking the prompt encoder for interactive segmentation in SAM. In future studies, we plan to explore the adaptation of SAM to interactive 3D medical segmentation.

\noindent\textbf{Acknowledgements.} This work was supported in part by grants from the National Natural Science Foundation of China under Grant No. 62306254, grants from the Hong Kong Innovation
and Technology Fund under Project ITS/030/21, grants from the Research
Grants Council of the Hong Kong Special Administrative Region, China (Project Reference Number: T45-401/22-N), and Project of Hetao Shenzhen-Hong Kong Science and Technology Innovation Cooperation Zone (HZQB-KCZYB-2020083).

\subsubsection{Disclosure of Interests.}
The authors have no competing interests to declare that are relevant to the content of this article.

\bibliographystyle{splncs04}
\bibliography{main}
\end{document}